\documentclass[conference]{IEEEtran}
\usepackage{comment}
\IEEEoverridecommandlockouts
\usepackage[nobreak,space,compress]{cite}
\usepackage[hidelinks]{hyperref}
\usepackage{url}
\usepackage{amsmath,amssymb,amsfonts}
\usepackage{algorithmic}
\usepackage{graphicx}
\usepackage{textcomp}
\usepackage{listings}
\usepackage{xcolor}
\usepackage{enumitem}
\usepackage{booktabs}
\usepackage[most]{tcolorbox}
\definecolor{DarkGreen}{RGB}{1,150,32}

\lstdefinestyle{PromptStyle}{
  basicstyle=\ttfamily\scriptsize,
  breaklines=true,
  frame=single,
  captionpos=b,
  rulecolor=\color{gray},
  escapeinside={(*@}{@*)},
  moredelim=**[is][\color{DarkGreen}\bfseries]{@d@}{@},
  moredelim=**[is][\color{blue}\bfseries]{@u@}{@},
  moredelim=**[is][\color{black}\bfseries]{@s@}{@}
}

\def\BibTeX{{\rm B\kern-.05em{\sc i\kern-.025em b}\kern-.08em
    T\kern-.1667em\lower.7ex\hbox{E}\kern-.125emX}}
\begin{document}

\title{Can LLMs Write CI? A Study on Automatic Generation of GitHub Actions Configurations
}

\author
{
  \IEEEauthorblockN{Taher A. Ghaleb and Dulina Rathnayake}
  \IEEEauthorblockA{Department of Computer Science,
  Trent University\\
  Peterborough, ON, Canada\\
  \{taherghaleb,dulinarathnayake\}@trentu.ca
  }
}

\maketitle

\begin{abstract}
Continuous Integration (CI) services, such as GitHub Actions, require developers to write YAML-based configurations, which can be tedious and error-prone.
Despite the increasing use of Large Language Models (LLMs) to automate software engineering tasks, their ability to generate CI configurations remains underexplored.
This paper presents a preliminary study evaluating six LLMs for generating GitHub Actions configurations from natural language descriptions. We assess three general-purpose foundation models (GPT-4o, Llama, and Gemma) and three code-pretrained models (GPT-4.1, Code~Llama, and CodeGemma). We also introduce the first labeled dataset of its kind, constructed from GitHub Actions documentation, pairing descriptions with corresponding best-practice YAML configurations.
Zero-shot prompting achieves up to $69\%$ similarity with the ground truth, with only $3\%$ perfect matches. Code-pretrained models slightly underperform compared to general-purpose ones in YAML-based CI tasks, revealing LLM limitations for CI configuration generation. Analyzing GPT-4o outputs reveals issues like missing or renamed steps, misinterpreted descriptions, and unnecessary additions that may affect structural and contextual correctness, indicating a gap between generation quality and the precision required for executable CI configurations.
Our research offers insights for improving LLM alignment with configuration languages and guiding future efforts on CI automation and tooling support.
\end{abstract}
\vspace{-7pt}

\vspace{-1.5pt}
\section{Introduction}
\vspace{-1.5pt}
Continuous Integration (CI) services are essential in modern software development for automating software building and testing~\cite{Fowler_CI}. GitHub Actions is a popular CI service due to its tight integration with GitHub and flexible workflow definitions. However, configuring CI involves writing YAML files that are complex both in syntax and semantics. Developers need to define job matrices, set up environments, and integrate third-party actions, often relying on diverse documentation~\cite{ghactions_workflow_syntax}. This can be time-consuming and error-prone, particularly for those who are new to CI concepts or YAML~\cite{ghactions_best_practices}.

Nowadays, Large Language Models (LLMs) are used for automating many software tasks like code generation, documentation, and bug fixing. However, generating CI's YAML configurations is challenging due to their non-executable nature as well as their strict indentation and domain-specific semantics~\cite{pujar2023automated}. The capability of LLMs, especially with natural language prompts, remains uncertain. Though much research studied CI configuration challenges~\cite{hilton2016usage,widder2019conceptual,zampetti2020empirical,vassallo2020configuration}, limited work addressed automatic generation~\cite{mastropaolo2024toward,zhang2024effectiveness}.

This paper presents a preliminary study evaluating how well LLMs can generate GitHub Actions configurations from natural language descriptions. We assess three general-purpose foundation models (GPT-4o, Llama3.1, and Gemma3) and three models pre-trained on source code (GPT-4.1, Code~Llama, and CodeGemma). To support our evaluation, and given the lack of existing datasets linking natural language to YAML configurations, we construct the first labeled dataset of its kind by curating \textit{Description–YAML} pairs from the official GitHub Actions documentation. Each pair includes a task-oriented description and a corresponding YAML snippet reflecting best practices. We use zero-shot prompting~\cite{kojima2022large} to simulate realistic developer usage without providing in-context examples. To measure the similarity between generated and reference configurations, we employ five metrics: \textit{Cosine Similarity}, \textit{Euclidean Distance}, \textit{Tree Edit Distance}, \textit{ROUGE-L}, and \textit{chrF}, capturing lexical overlap and structural correctness. This setup enables a detailed comparison of model outputs across both text-based and syntax-aware dimensions.

\vspace{0.8pt}
Overall, our findings highlight both the promise and limitations of LLMs for CI-related automation. They also indicate the need for better alignment between LLM architectures and configuration languages. In addition, our analysis reveals shortcomings in current CI documentation, which often lacks the clarity and specificity needed for accurate automation. Our work lays the foundation for future research on CI configuration synthesis and provides practical insights into the capabilities and boundaries of LLM-based automation tools.

\vspace{1pt}
This paper makes the following contributions.

\begin{itemize}
    \vspace{-1pt}
    \item Labeled dataset~\cite{our_replication_package} of $1,318$ YAML configurations with corresponding descriptions collected from official GitHub Actions documentation, supporting future research.
    
    \item Evaluation of six LLMs for generating CI configurations for GitHub Actions from natural language, using five metrics to assess lexical and structural similarity.
    
    \item Manual analysis of LLM-generated CI configurations, uncovering common inaccuracies and emphasizing the need for models better aligned with configuration tasks.
    \vspace{-2pt}
\end{itemize}

\vspace{1pt}
\noindent The rest of this paper is organized as follows. Section~\ref{background} covers background and related work on CI configuration and LLM applications. Section~\ref{approach} details dataset construction, selected LLMs, and evaluation setup. Section~\ref{evaluation} presents performance results and implications. Section~\ref{threats} discusses validity threats. Section~\ref{conclusion} concludes the paper.

\section{Background and Related Work}
\label{background}
\vspace{-3pt}
\subsection {Continuous Integration (CI)}
\vspace{-3pt}
Continuous Integration (CI) is a development practice where developers frequently merge code into a shared repository~\cite{Fowler_CI}. Each change triggers automated builds and tests to ensure stability and early identification of bugs. Quick feedback from CI improves collaboration and efficiency.

\vspace{-3pt}
\subsection{GitHub Actions Documentation}
\vspace{-3pt}
GitHub Actions is a CI service integrated in GitHub, in which workflows are defined using YAML files, specifying jobs to run and their triggers, such as pushes or pull requests. It contains many reusable actions for customization. However, configuring these files can be challenging without knowledge of YAML or GitHub Actions' rules.
GitHub Actions provides essential documentation for developers, such as guides, YAML references, and workflows. However, the abundance of features and updates can be overwhelming, leading to reliance on fragmented examples or forums, which complicates writing correct configurations, especially for CI newcomers.

\vspace{-3pt}
\subsection{Large Language Models (LLMs)}
\vspace{-3pt}
Large Language Models (LLMs), including GPT-4, Llama, and Gemma, are transformer-based models trained on extensive text and code datasets. They are adept at understanding and generating natural language, aiding software development tasks like code synthesis, explanation, and bug fixing. Recently, LLMs have been explored for DevOps tasks, such as generating configuration files from descriptions~\cite{rosa2023automatically,mehta2023automated}. However, creating CI configurations is challenging due to the strict syntax and structure of languages like YAML, and the specialized knowledge needed for valid CI workflows.

\vspace{-3pt}
\subsection{Studies on CI configurations}
\vspace{-3pt}
Prior research explored challenges in configuring and maintaining CI pipelines. Hilton et al.~\cite{hilton2016usage} surveyed over 400 developers and found that CI adoption is still limited in many open-source projects, often due to developers’ lack of experience with CI. Widder et al.~\cite{widder2019conceptual} and Zampetti et al.~\cite{zampetti2020empirical,ghaleb2025cicd} further documented recurring CI issues, including tool inconsistency and complex configuration. Similarly, Vassallo et al.~\cite{vassallo2020configuration} developed tooling support to detect configuration-related smells in deployment setups.
Poor CI configurations were reported to be associated with long build durations~\cite{ghaleb2019empirical,ghaleb2022studying} and build failures~\cite{ghaleb2019studying,ghaleb2022studying}.
Wurster et al.~\cite{rostami2023usage} explored CI service migration trends and observed growing adoption of GitHub Actions due to tighter platform integration and improved usability. These findings reinforce the relevance of automated CI configuration support, especially in light of growing configuration complexity and developer burden.

In the context of LLMs for CI, Mastropaolo et al.~\cite{mastropaolo2024toward} investigated workflow completion from partial YAMLs, while Zhang et al.~\cite{zhang2024effectiveness} evaluated LLMs using manually crafted scenarios, focusing on syntactic validity and execution feasibility. Despite these efforts, little work has examined the ability of LLMs to generate CI configurations directly from natural language descriptions, which our study addresses using a dataset derived from official documentation.

\section{Data Collection and Processing}
\label{approach}
\vspace{-2pt}
    \subsection{Data Construction}
        \vspace{-4pt}
        We constructed the first dataset of its kind that pairs linking natural language descriptions of Continuous Integration (CI) tasks to YAML-based GitHub Actions configurations. This dataset helps evaluate Large Language Models (LLMs) in generating accurate CI configurations from text descriptions. Data collection and processing were carefully designed to ensure quality, consistency, and relevance, using automated extraction from documentation and manual validation to build a diverse CI language task benchmark.
        
    \subsubsection{Data Sources}
        We developed a custom crawler using the Selenium automation framework~\footnote{\url{https://www.selenium.dev/documentation}} to systematically explore the official GitHub Actions documentation, starting with the main page~\footnote{\url{https://docs.github.com/en/actions}}, recursively navigating internal pages, and extracting instructional content. Non-informative elements such as sidebars, tables, footers, and navigation menus were removed to retain only meaningful text. The GitHub Actions documentation not only provides step-by-step guidance but also complete workflow examples and recommended best practices that developers are expected to follow.

    \subsubsection{Data Structure}
        Each example in the dataset consists of a structured pair that serves as a labeled reference to evaluate the ability of LLMs to generate CI configurations from natural language. Specifically:
        \begin{itemize}[leftmargin=0.4cm]
            \item \textbf{\textit{Description}}: A clear, task-specific natural language description.
            \item \textbf{\textit{Ground truth YAML}}: A verified YAML configuration that correctly implements the described task.
        \end{itemize}
        
        These pairs are in a CSV with standardized headers for easy automated processing and model output comparison.
                        
    \subsubsection{Preprocessing Steps}
        We preprocessed the dataset to maintain quality and consistency. First, YAML configurations were manually checked by the co-authors to ensure syntax and semantic accuracy with their descriptions. Then, we corrected indentation, removed extra whitespace, and fixed line breaks to prevent string evaluation errors. Finally, duplicate examples were removed, and entries with empty descriptions were excluded to ensure the dataset is meaningful and evaluable.
        Each entry showcases unique CI behavior, enabling dependable evaluation of LLM-generated configurations.
        
    \subsubsection{Dataset Summary}
        The dataset includes $1,318$ examples, each with a unique ID and a source URL from the GitHub Actions documentation. Each example has a description of the CI behavior and its YAML configuration. The complexity of descriptions and YAMLs varies, ranging from simple actions to complex multi-job workflows with features like matrices or conditional execution, offering a diverse benchmark for evaluating LLMs in CI tasks.
    
\section{Experimental Evaluation}
\label{evaluation}
\begin{figure*}[t]
      \vspace{-10pt}
      \centering
      \includegraphics[width=0.90\textwidth]{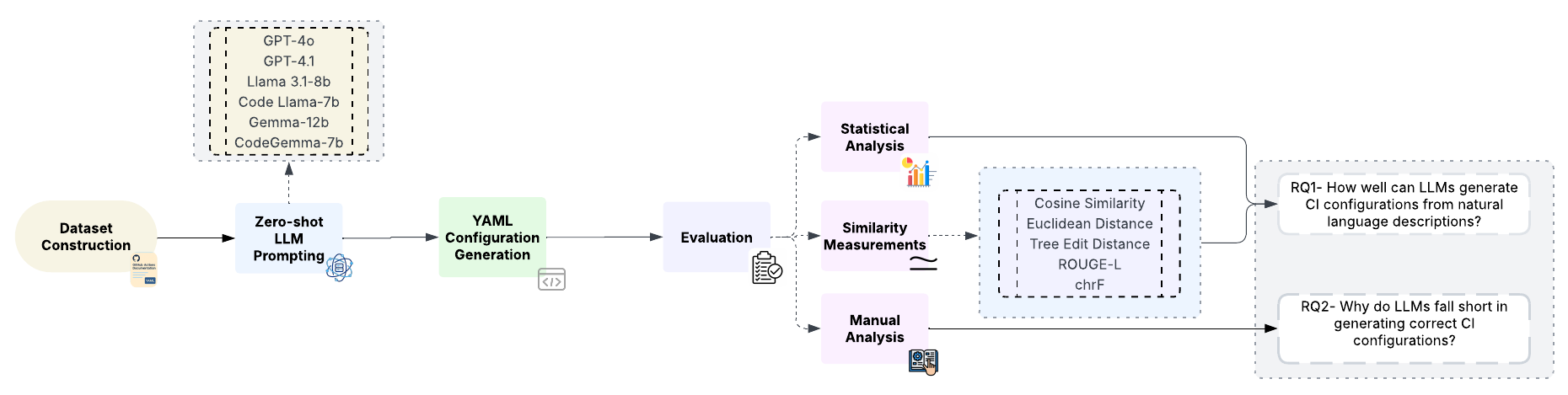}
      \vspace{-5pt}
      \caption{Overview of our study}
      \vspace{-10pt}
      \label{fig:study_overview}
    \end{figure*}

\vspace{-3pt}
This section details our experimental setup and results. Figure~\ref{fig:study_overview} provides an overview. We developed a labeled dataset of GitHub Actions descriptions paired with their corresponding YAML configurations, which we used to evaluate LLMs for YAML generation. Outputs were evaluated using similarity metrics and manual analysis. Statistical and structural comparisons address RQ1 on quality, and manual inspection finds common inaccuracies, addressing RQ2. Our replication package (data, scripts, and raw results) is publicly available on Figshare~\cite{our_replication_package}.

    \subsection{Selected Models}
    \label{sec:llms}
    \vspace{-3pt}
        We evaluated six publicly accessible LLMs, covering general-purpose and code-specialized capabilities. General-purpose models (GPT-4o, Llama3.1–8B, Gemma3–12B) are suited for natural language tasks, while code-pretrained models (GPT-4.1, Code~Llama–7B, CodeGemma–7B) focus on code synthesis and completion. 
        To stay within computational limits, open-source models of 7B to 12B parameters.
                
        \begin{itemize}[leftmargin=0.4cm]
            \item \textbf{\texttt{GPT-4o}}~\cite{openai2023gpt4o}: OpenAI's multimodal flagship model capable of processing and generating text, images, and audio in real time. It offers improved performance and reduced latency compared to its predecessors.
            
            \item \textbf{\texttt{GPT-4.1}}~\cite{openai2025gpt41}: An advanced transformer-based model by OpenAI, featuring major improvements in coding capabilities, instruction following, and long-context understanding.
            
            \item \textbf{\texttt{Llama3.1-8B}}~\cite{touvron2023Llama}: Part of Meta's Llama3.1 series, this 8-billion-parameter model is optimized for multilingual dialogue and instruction-following tasks.
            
            \item \textbf{\texttt{Code~Llama-7B}}~\cite{roziere2023codellama}: A code-specialized variant of Meta's Llama models, designed for general code synthesis and understanding, supporting tasks like code completion and infilling.
            
            \item \textbf{\texttt{Gemma3-12B}}~\cite{anil2024gemma}: A 12-billion-parameter model developed by Google, used in text generation and image understanding, such as question answering and summarization.
            
            \item \textbf{\texttt{CodeGemma-7B}}~\cite{zhao2024codegemma}: A lightweight code generation model built on Google's Gemma architecture, specializing in code completion and generation tasks.
        \end{itemize}
                
    \vspace{-2pt}
    \subsection{Evaluation Setup}
        \vspace{-3pt}
        We used a standardized prompt template (Listing~\ref{lst:prompt-template}) to ensure all models return only YAML configurations, isolating their ability to translate task descriptions into executable CI configurations.
        We set the \texttt{temperature} parameter to \texttt{0} across all models to reduce randomness and enforce deterministic behavior. GPT-4o and GPT-4.1 were accessed using the OpenAI API\footnote{\url{https://platform.openai.com}}, while the open-source models (Llama, Code~Llama, Gemma, and CodeGemma) were run locally using the Ollama framework\footnote{\url{https://ollama.com}}. We used the maximum context length supported by each model to avoid input truncation.
        The generated YAML configurations were extracted and appended to the dataset for further evaluation.

\begin{figure}
\begin{lstlisting}[
  style=PromptStyle,
  label={lst:prompt-template},
  caption={Prompt template used for our LLMs}
]
@u@System:@
@s@You are a highly skilled software developer and DevOps
engineer.@

@u@User:@
@s@Given the description below, generate a minimal and single
valid GitHub Actions YAML configuration. Do not provide any
reasoning. Only output the single YAML configuration, with
no comments, markdown, or extra formatting. The YAML must
match the described task exactly. Do not add any unneeded
directives unless told.
@
@d@{Description}@
\end{lstlisting}
\vspace{-18pt}
\end{figure}

    \vspace{-5pt}
    \subsection{Evaluation Metrics}
    \vspace{-2pt}
    \label{sec:metrics}
        To assess the similarity between LLM-generated configurations and the ground truth, we employed five metrics that capture both lexical and structural differences. These are crucial for identifying issues in configurations where minor variations can lead to failures in CI workflows.
        
        \begin{itemize}[leftmargin=0.4cm]
            \item \textbf{\textit{Cosine Similarity}} measures the similarity of reference and generated configurations using the cosine similarity of their TF-IDF vectors to capture the lexical token overlap~\cite{salton1983introduction}.
            
            \item \textbf{Tree Edit Similarity} is the inverse of the tree edit distance between parsed YAML structures, calculated by the minimum node operations (insertions, deletions, substitutions) required to transform one configuration to another~\cite{zhang1989simple}.
            
            \item \textbf{Euclidean Similarity} is the inverse of the Euclidean distance between TF-IDF vectors, accounting for both direction and magnitude, complementing cosine similarity. Though not often cited alone, Euclidean distance is essential in vector space models and unsupervised learning~\cite{manning2008introduction}.
        
            \item \textbf{\textit{ROUGE-L}} examines the longest common subsequence between two sequences, focusing on fluency and order. It is often used in summarization evaluation~\cite{lin2004rouge}.
            
            \item \textbf{\textit{chrF}} calculates character n-gram F-scores, detecting word-level changes and capturing fine-grained variations~\cite{popovic2015chrf}.
                        
        \end{itemize}

    \subsection{Evaluation Results}   
            \vspace{-2pt}
            \subsubsection{\textbf{RQ1: How well can LLMs generate CI configurations from natural language descriptions?}}~
    
                \vspace{2pt}
                \noindent\textbf{[Motivation]}~
                    LLMs are widely used for code generation, but their ability to generate correct CI configurations (for GitHub Actions, in particular) remains unclear. These workflows are sensitive to structure and syntax, and small errors can lead to failed automation. This question evaluates how well current LLMs can generate valid GitHub Actions YAML configurations from natural language descriptions.
                    
                \vspace{3pt}
                \noindent\textbf{[Approach]}~
                We evaluate six LLMs: three general-purpose and three code-pretrained (see Section~\ref{sec:llms}) using a standardized zero-shot prompt~\cite{brown2020language} to generate GitHub Actions configurations from 1,318 natural language descriptions. The outputs are compared against ground truth using five similarity metrics (see Section~\ref{sec:metrics}). We used Wilcoxon signed-rank~\cite{wilcoxon1945individual} to test for significant differences between similarity scores of the five models. To test whether description length (i.e., word count) is associated with LLM performance, we computed Spearman correlations~\cite{spearman1904} between word count and the five similarity metrics of GPT-4o. We interpret effect size using Spearman’s $\rho$~\cite{cohen1988statistical}. We further applied Holm–Bonferroni correction~\cite{holm1979simple} to adjust for multiple comparisons.

\begin{figure*}
                        \centering
                        \vspace{-11pt}\includegraphics[width=0.95\textwidth]{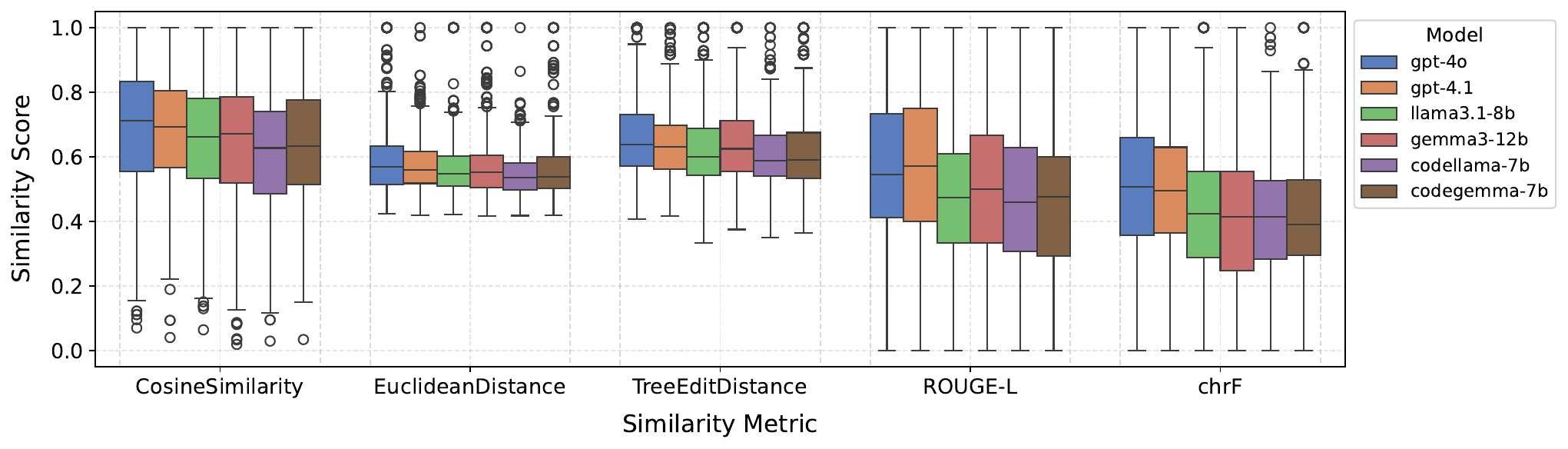}
                        \vspace{-8pt}
                        \caption{Boxplots showing the distribution of similarity scores across six LLMs using five evaluation metrics in generating YAML configurations}
                        \vspace{-10pt}
                    \label{fig:llms_similarity_results}
                    \end{figure*}

                \vspace{3pt}
                \noindent\textbf{[Results]}~
                    Figure~\ref{fig:llms_similarity_results} shows boxplots of similarity scores for the six~LLMs using five evaluation metrics, highlighting differences in lexical and structural alignment with the reference YAML configurations.
                    We observe that lexical-based metrics (Cosine Similarity and ROUGE-L) show wider performance gaps across models, indicating greater variation in how LLMs interpret and phrase task descriptions. In contrast, structural metrics (Tree Edit Similarity and Euclidean Similarity) are more uniformly distributed, suggesting that even weaker models often preserve YAML structure to some extent.

                    \vspace{3pt}
                    \noindent\textbf{Model-wise Performance Comparison.}
                    GPT-4o outperformed all other models across most metrics. It achieved the highest average Cosine Similarity (0.69) and Euclidean Similarity (0.59), demonstrating strong lexical and structural alignment with the ground truth, with a perfect match in only $3\%$ of cases. GPT-4.1 was slightly lower than GPT-4o on all other metrics.
                    Code-pretrained models (i.e., Code~Llama and CodeGemma) consistently ranked in the bottom tier. Their average Cosine Similarity scores (0.61 and 0.64) were significantly lower than GPT-4o’s ($p$-$value<0.001$) and also lower than their general-purpose counterparts, Llama (0.65) and Gemma3 (0.65). Their Euclidean Similarity scores (0.55 and 0.56) were also the lowest among all models compared to GPT-4o ($p$-$value<0.001$). These results challenge the assumption that code-centric pretraining inherently improves YAML generation.

                    \vspace{3pt}
                    \noindent\textbf{Structural Similarity.}
                    Tree Edit Similarity showed GPT-4o (0.66) slightly ahead of other models, indicating better structural alignment with reference configurations. Surprisingly, Llama3.1 (0.63) and Gemma3 (0.65) performed better than Code~Llama (0.61) and CodeGemma (0.62), despite lacking code-specific pretraining. Euclidean Similarity confirmed this trend, where GPT-4o achieved the highest similarity (0.59), while CodeGemma and Code~Llama achieved the lowest (both at 0.56).

                    \vspace{4pt}
                    \noindent\textbf{Impact of Description Length.}
                    Our correlation analysis revealed that description length is statistically correlated with GPT-4o's output quality, but with small effect sizes. Cosine Similarity and Euclidean Distance showed the strongest correlations ($\rho=0.198$, $p$-$value<0.001$), followed by ROUGE-L ($\rho=0.187$, $p$-$value<0.001$), while Tree Edit Distance had a weaker but still significant correlation ($\rho=0.118$, $p$-$value<0.001$). These findings suggest that longer descriptions may slightly improve lexical and structural alignment.

            \subsubsection{\textbf{RQ2: Why do LLMs fall short in generating correct CI configurations?}}~
    
                \vspace{3pt}
                \noindent\textbf{[Motivation]}
                Automated metrics assess similarity yet fail to explain the quality of LLMs' output. Small configuration mistakes can misrepresent intent, highlighting the need for understanding such mistakes. Qualitative analysis uncovers weaknesses unnoticed by numerical evaluations, crucial for improving prompting, training, and CI/LLM integration.
                
                \vspace{3pt}
                \noindent\textbf{[Approach]}
                We manually analyzed a random statistical sample of $90$ YAML configurations generated by GPT-4o (with a $95\%$ confidence level and $\pm10\%$ margin of error). The goal was to determine if inaccuracies arose from the description, the LLM's output, or both by comparing the generated YAML to its ground truth and input description. To this end, the two co-authors, both experienced in software engineering and GitHub Actions, collaboratively analyzed the sample in discussion meetings using card sorting~\cite{spencer2009card}.

                \vspace{3pt}
                \noindent\textbf{[Results]}
                We found that LLMs often struggle not with syntactically invalid YAML, but with interpreting the specification's intent and details. GPT-4o often fails when crucial information is not clearly highlighted in descriptions. For example, when general usage of \texttt{workflow\_call.inputs} is referenced, it defaults to generic fields like \texttt{example\_input} instead of specific ones like username, leading to incorrect setups, reflecting a tendency to generalize with vague input.
                Our analysis revealed an oversight of clearly stated elements, such as caching, prioritizing core concepts over essential directives. The model simply ignored such elements, assuming they are optional, leading to missing fields necessary for accuracy.
                
                The model often makes speculative additions, or hallucinations, not present in the description. For instance, it might assume a checkout step with \texttt{actions/checkout@v4} or include fields like \textit{environment.url}, even if not specified. These additions, though reflecting common patterns, misalign with the intended description.
                Naming inconsistencies and reformatting, like rewording or reordering job or step names, are also common. Though usually harmless due to the subjective nature of naming, they can reduce traceability and hinder comparison with reference configurations during evaluation.  
                
                These patterns highlight recurring challenges: distinguishing optional from essential components, avoiding excessive YAML completion, and maintaining precision over readability. Creating effective CI configurations demands deep understanding of the task context and constraints, beyond surface-level fluency.\vspace{-3pt}
       
        \vspace{5pt}
        \subsubsection{Discussion}~

        \vspace{2pt}
        \noindent\textbf{Need for CI-configuration oriented LLMs.}  
            Our RQ1 findings reveal that GPT-4o scores higher in lexical and structural similarity compared to Code~Llama or CodeGemma, indicating that broad language understanding might be more crucial than code-centric pretraining for YAML synthesis. As part of our future work, we plan to assess more LLMs of different sizes to explore the impact of scale and architecture on CI configuration quality.
                
        \vspace{3pt}
        \noindent\textbf{Precise CI descriptions improve quality.}  
            RQ2 revealed that many well-structured configurations failed due to missing steps, wrong parameters, or added elements, highlighting the need for domain-aware validation beyond format alignment. Though weak correlations were found between description length and output quality, our manual analysis showed that vague inputs often misled the model. Improving CI documentation and urging developers to give clear, task-specific prompts are crucial for reliable LLM-assisted configuration.
            
        \vspace{3pt}
        \noindent\textbf{Prompting must reflect YAML semantics.}  
            Some inaccuracies in RQ2 were related to hallucinated defaults or omitted mandatory keys by LLMs. Fine-tuning LLMs~\cite{raffel2020exploring} with YAML-specific constraints and schema checks can help reduce such issues, which we plan to explore in future work. In addition, we plan to use Retrieval-Augmented Generation (RAG)~\cite{lewis2020retrieval} to reference CI documentation during the generation of CI configurations, which can further improve factual accuracy and reduce unsupported assumptions.
        
\section{Threats to Validity}
\label{threats}
\vspace{-2pt}
    \noindent\textbf{Construct Validity.}
        The similarity metrics used in our study aim to approximate lexical and structural alignment of ground truth and LLM-generated configurations, but may not fully reflect practical executability or correctness. The qualitative analysis was performed to uncover inaccuracies in generated outputs, but it might not capture all edge cases. Also, slight rewording or renaming (e.g., job names) was accepted unless it altered behavior, but these boundary decisions could influence the reported frequencies. Another potential threat is data leakage, as LLMs may have been exposed to the GitHub Actions documentation used in our dataset. While we cannot verify this, the limited performance (i.e., only 3\% perfect matches) suggests that such models still struggle with reliably generating correct configurations. 

    \vspace{4pt}
    \noindent\textbf{Internal Validity.}
        Our manual evaluation of GPT-4o-generated YAML configurations may be subject to reviewer bias. To mitigate this, the two co-authors with domain expertise conducted all assessments collaboratively through discussion meetings, ensuring agreement on every annotation. Still, subjective judgments could influence error categorization. Moreover, the sample size (90 configurations), while statistically grounded, may not fully capture the diversity of CI configuration scenarios in practice, reflecting the preliminary nature of this study.
        
    \vspace{4pt}
    \noindent\textbf{External Validity.}
        We focused on GitHub Actions configurations, which may limit the generalizability of our findings to other CI services (e.g., Travis CI or CircleCI). In addition, we evaluated a single model (GPT-4o) for RQ2, which is the top-performing model. However, other LLMs may exhibit different patterns of inaccuracies. Moreover, we did not execute or validate the correctness of the ground truth YAML configurations, as they were directly extracted from the official GitHub Actions documentation. While many of them are partial, we assumed the documentation to be authoritative and accurate.

\vspace{-2pt}
\section{Conclusion}
\label{conclusion}
\vspace{-3pt}
    This paper presents a preliminary study evaluating six language models in generating GitHub Actions CI configurations from natural language. We created the first labeled dataset of CI configuration errors for better analysis. GPT-4o outperformed both general-purpose and code-pretrained models in metrics. Notably, code-focused models (Code~Llama and CodeGemma) did not consistently outperform general-purpose models (Llama and Gemma), questioning code-centric training's efficacy. Our manual analysis revealed common issues with the generated configurations due to vague prompts or model misassumptions. Our study highlights the need for clearer prompting and CI-aware LLMs that are better aligned with YAML syntax and CI processes to improve prompt-to-configuration accuracy and automation efficiency, which we aim to explore in future work.

\section*{Acknowledgment}
This research was supported by the Natural Sciences and Engineering Research Council (NSERC) Discovery Grant [RGPIN-2025-05897] and Trent University Knowledge Mobilization Research Grant. The experiments conducted in this paper were enabled in part by support provided by the Digital Research Alliance of Canada (\url{https://alliancecan.ca}).

\clearpage
\bibliographystyle{IEEEtran}
\bibliography{paper}

\begin{thebibliography}{10}
\providecommand{\url}[1]{#1}
\csname url@samestyle\endcsname
\providecommand{\newblock}{\relax}
\providecommand{\bibinfo}[2]{#2}
\providecommand{\BIBentrySTDinterwordspacing}{\spaceskip=0pt\relax}
\providecommand{\BIBentryALTinterwordstretchfactor}{4}
\providecommand{\BIBentryALTinterwordspacing}{\spaceskip=\fontdimen2\font plus
\BIBentryALTinterwordstretchfactor\fontdimen3\font minus \fontdimen4\font\relax}
\providecommand{\BIBforeignlanguage}[2]{{%
\expandafter\ifx\csname l@#1\endcsname\relax
\typeout{** WARNING: IEEEtran.bst: No hyphenation pattern has been}%
\typeout{** loaded for the language `#1'. Using the pattern for}%
\typeout{** the default language instead.}%
\else
\language=\csname l@#1\endcsname
\fi
#2}}
\providecommand{\BIBdecl}{\relax}
\BIBdecl

\bibitem{Fowler_CI}
M.~Fowler, ``Continuous {Integration},'' \url{https://martinfowler.com/articles/originalContinuousIntegration.html}, accessed: 2021-09-23.

\bibitem{ghactions_workflow_syntax}
``Workflow syntax for {GitHub Actions},'' \url{https://docs.github.com/en/actions/writing-workflows}, Accessed May 2025.

\bibitem{ghactions_best_practices}
``{GitHub Actions} best practices,'' \url{https://docs.github.com/en/actions/use-cases-and-examples}, Accessed May 2025.

\bibitem{pujar2023automated}
S.~Pujar, L.~Buratti, X.~Guo, N.~Dupuis, B.~Lewis, S.~Suneja, A.~Sood, G.~Nalawade, M.~Jones, A.~Morari, and R.~Puri, ``Automated code generation for information technology tasks in {YAML} through large language models,'' \emph{arXiv preprint arXiv:2305.02783}, 2023.

\bibitem{hilton2016usage}
M.~Hilton, T.~Tunnell, K.~Huang, D.~Marinov, and D.~Dig, ``Usage, costs, and benefits of continuous integration in open-source projects,'' in \emph{Proceedings of the 31st IEEE/ACM international conference on automated software engineering}, 2016, pp. 426--437.

\bibitem{widder2019conceptual}
D.~G. Widder, M.~Hilton, C.~K{\"a}stner, and B.~Vasilescu, ``A conceptual replication of continuous integration pain points in the context of {Travis CI},'' in \emph{Proceedings of the 2019 27th acm joint meeting on european software engineering conference and symposium on the foundations of software engineering}, 2019, pp. 647--658.

\bibitem{zampetti2020empirical}
F.~Zampetti, C.~Vassallo, S.~Panichella, G.~Canfora, H.~Gall, and M.~Di~Penta, ``An empirical characterization of bad practices in continuous integration,'' \emph{Empirical Software Engineering}, vol.~25, pp. 1095--1135, 2020.

\bibitem{vassallo2020configuration}
C.~Vassallo, S.~Proksch, A.~Jancso, H.~C. Gall, and M.~Di~Penta, ``Configuration smells in continuous delivery pipelines: a linter and a six-month study on {GitLab},'' in \emph{Proceedings of the 28th ACM Joint Meeting on European Software Engineering Conference and Symposium on the Foundations of Software Engineering}, 2020, pp. 327--337.

\bibitem{mastropaolo2024toward}
A.~Mastropaolo, F.~Zampetti, G.~Bavota, and M.~Di~Penta, ``Toward automatically completing {GitHub} workflows,'' in \emph{Proceedings of the 46th IEEE/ACM International Conference on Software Engineering}, 2024, pp. 1--12.

\bibitem{zhang2024effectiveness}
X.~Zhang, S.~Muralee, S.~Cherupattamoolayil, and A.~Machiry, ``On the effectiveness of large language models for {GitHub} workflows,'' in \emph{Proceedings of the 19th International Conference on Availability, Reliability and Security}, 2024, pp. 1--14.

\bibitem{kojima2022large}
T.~Kojima, S.~S. Gu, M.~Reid, Y.~Matsuo, and Y.~Iwasawa, ``Large language models are zero-shot reasoners,'' \emph{Advances in neural information processing systems}, vol.~35, pp. 22\,199--22\,213, 2022.

\bibitem{our_replication_package}
T.~A. Ghaleb and D.~Rathnayake, ``Can {LLMs} write {CI}? a study on automatic generation of github actions configurations (replication package),'' \url{https://github.com/Taher-Ghaleb/ICSME25-LLM4CI}, 2025.

\bibitem{rosa2023automatically}
G.~Rosa, A.~Mastropaolo, S.~Scalabrino, G.~Bavota, and R.~Oliveto, ``Automatically generating {Dockerfiles} via deep learning: Challenges and promises,'' \emph{arXiv preprint arXiv:2303.15990}, 2023.

\bibitem{mehta2023automated}
D.~Mehta, K.~Rawool, S.~Gujar, and B.~Xu, ``Automated {DevOps} pipeline generation for code repositories using large language models,'' \emph{arXiv preprint arXiv:2312.13225}, 2023.

\bibitem{ghaleb2025cicd}
T.~Ghaleb, O.~Abduljalil, and S.~Hassan, ``{CI/CD} configuration practices in open-source {Android} apps: An empirical study,'' \emph{ACM Transactions on Software Engineering and Methodology}, 2025.

\bibitem{ghaleb2019empirical}
T.~A. Ghaleb, D.~A. Da~Costa, and Y.~Zou, ``An empirical study of the long duration of continuous integration builds,'' \emph{Empirical Software Engineering}, vol.~24, no.~4, pp. 2102--2139, 2019.

\bibitem{ghaleb2022studying}
T.~A. Ghaleb, S.~Hassan, and Y.~Zou, ``Studying the interplay between the durations and breakages of continuous integration builds,'' \emph{IEEE Transactions on Software Engineering}, vol.~49, no.~4, pp. 2476--2497, 2022.

\bibitem{ghaleb2019studying}
T.~A. Ghaleb, D.~A. Da~Costa, Y.~Zou, and A.~E. Hassan, ``Studying the impact of noises in build breakage data,'' \emph{IEEE Transactions on Software Engineering}, vol.~47, no.~9, pp. 1998--2011, 2019.

\bibitem{rostami2023usage}
P.~Rostami~Mazrae, T.~Mens, M.~Golzadeh, and A.~Decan, ``On the usage, co-usage and migration of {CI/CD} tools: A qualitative analysis,'' \emph{Empirical Software Engineering}, vol.~28, no.~2, p.~52, 2023.

\bibitem{openai2023gpt4o}
\BIBentryALTinterwordspacing
O.~AI, ``{GPT-4} technical report,'' \emph{arXiv preprint arXiv:2303.08774}, 2023. [Online]. Available: \url{https://arxiv.org/abs/2303.08774}
\BIBentrySTDinterwordspacing

\bibitem{openai2025gpt41}
\BIBentryALTinterwordspacing
OpenAI, ``Introducing {GPT-4.1} in the {API},'' 2025, accessed: 2025-06-04. [Online]. Available: \url{https://openai.com/index/gpt-4-1/}
\BIBentrySTDinterwordspacing

\bibitem{touvron2023Llama}
\BIBentryALTinterwordspacing
H.~Touvron, T.~Lavril, G.~Izacard, X.~Martinet, M.-A. Lachaux, T.~Lacroix, B.~Rozière, N.~Goyal, E.~Hambro, F.~Azhar, A.~Rodriguez, A.~Joulin, E.~Grave, and G.~Lample, ``{LLaMA}: Open and efficient foundation language models,'' \emph{arXiv preprint arXiv:2302.13971}, 2023. [Online]. Available: \url{https://arxiv.org/abs/2302.13971}
\BIBentrySTDinterwordspacing

\bibitem{roziere2023codellama}
\BIBentryALTinterwordspacing
B.~Rozière, J.~Gehring, S.~Bekman, M.~Ott, T.~Scialom, S.~Edunov, and G.~Lample, ``Code llama: Open foundation models for code,'' \emph{arXiv preprint arXiv:2308.12950}, 2023. [Online]. Available: \url{https://arxiv.org/abs/2308.12950}
\BIBentrySTDinterwordspacing

\bibitem{anil2024gemma}
\BIBentryALTinterwordspacing
R.~Anil \emph{et~al.}, ``Gemma 3 technical report,'' \emph{arXiv preprint arXiv:2503.19786}, 2024. [Online]. Available: \url{https://arxiv.org/abs/2503.19786}
\BIBentrySTDinterwordspacing

\bibitem{zhao2024codegemma}
\BIBentryALTinterwordspacing
H.~Zhao, J.~Hui, J.~Howland, N.~Nguyen, S.~Zuo, A.~Hu, C.~A. Choquette-Choo, J.~Shen, J.~Kelley, K.~Bansal, L.~Vilnis, M.~Wirth, P.~Michel, P.~Choy, P.~Joshi, R.~Kumar, S.~Hashmi, S.~Agrawal, Z.~Gong, J.~Fine, T.~Warkentin, A.~J. Hartman, B.~Ni, K.~Korevec, K.~Schaefer, and S.~Huffman, ``Codegemma: Open code models based on gemma,'' \emph{arXiv preprint arXiv:2406.11409}, 2024. [Online]. Available: \url{https://arxiv.org/abs/2406.11409}
\BIBentrySTDinterwordspacing

\bibitem{salton1983introduction}
G.~Salton and M.~J. McGill, \emph{Introduction to modern information retrieval}.\hskip 1em plus 0.5em minus 0.4em\relax McGraw-Hill, 1983.

\bibitem{zhang1989simple}
K.~Zhang and D.~Shasha, ``Simple fast algorithms for the editing distance between trees and related problems,'' \emph{SIAM Journal on Computing}, vol.~18, no.~6, pp. 1245--1262, 1989.

\bibitem{manning2008introduction}
C.~D. Manning, P.~Raghavan, and H.~Schütze, \emph{Introduction to Information Retrieval}.\hskip 1em plus 0.5em minus 0.4em\relax Cambridge University Press, 2008.

\bibitem{lin2004rouge}
C.-Y. Lin, ``{ROUGE}: A package for automatic evaluation of summaries,'' in \emph{Text summarization branches out}, 2004, pp. 74--81.

\bibitem{popovic2015chrf}
M.~Popovi{\'c}, ``{chrF}: character n-gram {F-score} for automatic {MT} evaluation,'' in \emph{Proceedings of the tenth workshop on statistical machine translation}, 2015, pp. 392--395.

\bibitem{brown2020language}
T.~B. Brown, B.~Mann, N.~Ryder, M.~Subbiah, J.~Kaplan, P.~Dhariwal, A.~Neelakantan, P.~Shyam, G.~Sastry, A.~Askell, S.~Agarwal, A.~Herbert-Voss, G.~Krueger, T.~Henighan, R.~Child, A.~Ramesh, D.~M. Ziegler, J.~Wu, C.~Winter, C.~Hesse, M.~Chen, E.~Sigler, M.~Litwin, S.~Gray, B.~Chess, J.~Clark, C.~Berner, S.~McCandlish, A.~Radford, I.~Sutskever, and D.~Amodei, ``Language models are few-shot learners,'' in \emph{Advances in Neural Information Processing Systems}, vol.~33, 2020.

\bibitem{wilcoxon1945individual}
F.~Wilcoxon, ``Individual comparisons by ranking methods,'' \emph{Biometrics Bulletin}, vol.~1, no.~6, pp. 80--83, 1945.

\bibitem{spearman1904}
C.~Spearman, ``The proof and measurement of association between two things,'' \emph{The American Journal of Psychology}, vol.~15, no.~1, pp. 72--101, 1904.

\bibitem{cohen1988statistical}
J.~Cohen, \emph{Statistical Power Analysis for the Behavioral Sciences}.\hskip 1em plus 0.5em minus 0.4em\relax Routledge, 1988.

\bibitem{holm1979simple}
S.~Holm, ``A simple sequentially rejective multiple test procedure,'' \emph{Scandinavian Journal of Statistics}, vol.~6, no.~2, pp. 65--70, 1979.

\bibitem{spencer2009card}
D.~Spencer, \emph{Card sorting: Designing usable categories}.\hskip 1em plus 0.5em minus 0.4em\relax Rosenfeld Media, 2009.

\bibitem{raffel2020exploring}
C.~Raffel, N.~Shazeer, A.~Roberts, K.~Lee, S.~Narang, M.~Matena, Y.~Zhou, W.~Li, and P.~J. Liu, ``Exploring the limits of transfer learning with a unified text-to-text transformer,'' \emph{Journal of Machine Learning Research}, vol.~21, no. 140, pp. 1--67, 2020.

\bibitem{lewis2020retrieval}
P.~Lewis, E.~Perez, A.~Piktus, F.~Petroni, V.~Karpukhin, N.~Goyal, I.~Kulikov, A.~Fan, V.~Chaudhary, F.~Guzman \emph{et~al.}, ``Retrieval-augmented generation for knowledge-intensive {NLP} tasks,'' in \emph{Advances in Neural Information Processing Systems}, vol.~33, 2020, pp. 9459--9474.

\end{thebibliography}

\end{document}